\documentclass[12pt]{iopart}
\pdfoutput=1
\usepackage{graphicx,graphics}

\begin{document}

\title[Charge carrier density noise in graphene]{Charge carrier density noise in graphene: effect of localized/delocalized traps}
\author{Francesco M.D. Pellegrino,$^{1\,,2}$, Giuseppe Falci,$^{1\,,2\,,3}$ and
Elisabetta Paladino$^{1\,,2\,,3}$}
\address{$^1$ Dipartimento di Fisica e Astronomia ``Ettore Majorana'', \\Universit\`a di Catania, Via S. Sofia 64, I-95123 Catania,~Italy}
\address{$^2$  INFN, Sez.~Catania, I-95123 Catania,~Italy}
\address{$^3$ CNR-IMM, Via S. Sofia 64, I-95123 Catania,~Italy}
\ead{epaladino@dmfci.unict.it}
\begin{abstract}
Graphene-based devices show $1/f$ low-frequency noise in several electronic transport properties, such as
mobility and charge carrier concentration.
The recent outburst of experimental studies on graphene-based devices integrated  
into circuit quantum electrodynamics systems has rekindled the interest in low frequency charge noise. 
We investigate charge carrier density noise in graphene within the McWorther model
where  noise is induced by electron traps in the substrate.
We focus on the large doping regime and introduce a simple modelization of the effect of localized/delocalized 
traps in terms of single/double spin occupancy of trap states.
We find that in both cases the charge carrier spectrum of graphene obeys the $1/f^\alpha$ power-law behavior where
$\alpha$ is very close to the unity, and  for each case we evaluate the deviation $\beta=1-\alpha$. 
The amplitude of the noise is found to depend on the trap energy distribution and on 
temperature.  Single/double spin occupancy  of trap states 
influences the temperature dependence of noise amplitude only to second order. 
\end{abstract}
\maketitle

\section{Introduction}

The class of devices based on graphene encapsulated in hexagonal boron-nitride (hBN) shows remarkable transport properties.
The encapsulation makes graphene unaffected by ambient atmosphere, moreover  hBN weakly influences the transport properties of graphene thanks to ultra-cleanness, 
large band gap and excellent lattices matching.
Based on this methodology, it has been possible to fabricate devices  which exhibit pronounced ballistic 
transport features over micrometer scale for a wide range of carrier concentrations and up to quite high temperatures 
($T \sim 50~{\rm K}$)~\cite{dean_natnano_2010,mayorov_nanolett_2011,wang_science_2013}.
Moreover, by using devices based on graphene encapsulated in hBN,
thanks to the weak electron-phonon coupling, at temperatures large enough to soften the Pauli blocking, the 
electron system can  exhibit measurable phenomena characteristic of hydrodynamic transport,
such as the negative nonlocal resistance~\cite{bandurin_science_2016,torre_prb_2015,fmdp_prb_2016}, 
the superballistic transport through a constriction~\cite{kumar_natphys_2017}, 
and the appearance of the Hall viscosity~\cite{berdyugin_science_2019,fmdp_prb_2017}.
Combination of these high-quality graphene samples with superconductors via clean interfaces has led to ballistic 
transport of Cooper pairs over micron scale lengths~\cite{benshalom_natph_2015, calado_natnano_2015, english_prb_2016}.
Recently, planar graphene Josephson junctions, formed by arranging two superconducting electrodes in close proximity 
of a graphene layer, have been investigated~\cite{borzenets_prl_2016, nanda_nano_2017, park_prl_2018}.
In this context, there has been an outburst of experimental studies on devices based on graphene encapsulated 
in hBN integrated  into circuit quantum electrodynamics systems~\cite{kroll_natcomm_2018,schmidt_natcomm_2018,wang_natnanotech_2018}.
Understanding charge noise in graphene based devices is a relevant issue in different transport regimes and in a wide temperature range. 
Charge carrier noise is one of the possible sources of $1/f$ noise observed in single- and multi-layer graphene and graphene based devices~\cite{balandin_natnano_2013}.
At low temperatures the charge density noise is expected to influence the decoherence  of graphene-based "gatemon", 
similarly to the effect on conventional superconducting qubits~
~\cite{larsen_prl_2015, paladino_rmp_2014,chiarello_njp_2012,pellegrino_proceedings_2019}.
Moreover, it can assist the occurrence of electronic preturbulence phenomena in hydrodynamic transport regime at room temperature~\cite{gabbana_prl_2018}.

In this manuscript, we investigate charge carrier density noise in graphene within the McWorther model
commonly used to describe  noise induced by electron traps in oxide substrates,  
unavoidably present because of  chemical or structural disorder~\cite{hooge_rps_1981,balandin_natnano_2013}.
 Here we do not specify the graphene substrate, most used being SiO$_2$ or hBN. 
The microscopic characteristics of the electron traps depend on the nature of the substrate.
 Detection of the power spectral density of the fluctuations of charge carrier density graphene on Si/SiO$_2$ 
has been recently reported by Hall voltage measurements~\cite{lu_prb_2014} and described within the tunnel/trap
model in \cite{Sun_JLT_2013}. Charge traps act as independent generation-recombination centers, which are described 
as random telegraph processes.
Here, we focus on large doping regime ($n\sim 10^{12}~{\rm cm}^{-2}$) and we introduce  
a simple modelization of the effect of localized/delocalized 
traps in terms of single/double spin occupancy of trap states.
For spatially localized traps we assume that 
the Coulomb repulsion between electrons forbids the spin double occupation of trap states. 
The simultaneous presence of two electrons with opposite spin $\hat{z}$-projection in the trap has in fact a large
energetic cost~\cite{grosso_book}. 
For delocalized traps~\cite{brews_jap_1972} the electron-electron repulsion is assumed to be negligible, allowing
for spin degeneracy.
We find that in both cases the charge carrier spectrum of graphene obeys the $1/f^\alpha$ power-law behavior where
$\alpha$ is very close to the unity, and for each case we evaluate the deviation $\beta=1-\alpha$. 
The amplitude of the noise is found to depend on the trap energy distribution and on 
temperature.  Single/double spin occupancy  of trap states (corresponding to localized/delocalized traps) 
influences the temperature dependence of noise amplitude only to second order.

\section{Model}

In this Section,  we introduce a simple model relating charge carrier density in graphene to 
charge traps in the substrate.
We refer to a simple device formed by a monolayer graphene on a substrate hosting electron traps
placed on top of a metal gate. 
The metal-substrate interface is at $z=0$, the graphene layer is treated as a two-dimensional system at $z=d$, and
the substrate width $d$ is much smaller than both longitudinal dimensions $L_x$ and $L_y$, along $\hat{x}$ and $\hat{y}$  directions. 
The gate voltage drop $V_{\rm G}$ between the metal gate and the graphene layer is fixed and
charge carriers tunneling between the graphene electron channel and substrate
trap states induces a fluctuating voltage, $V_{\rm T} (t)$.
The gate voltage is given by
\begin{equation}\label{eq:Vg}
 V_{\rm G} =   V_{\rm T} (t)+\frac{e  n(t)}{C_{\rm g}} + \frac{\mu(t)}{e}  +\frac{W_{\rm f}}{e}~,
 \end{equation}
where $e  n(t)/C_{\rm g}$ is the electrostatic contribution due to the charge carriers density 
$n(t)$ in graphene layer, with 
the geometric capacitance $C_{\rm g}=\varepsilon_r/(4 \pi d)$ expressed in terms of the dielectric constant of the substrate.
The chemical potential of graphene,  $\mu(t)$, takes into account quantum corrections to the classical geometric capacitance~\cite{fernandez_prb_2007},
and $W_{\rm f}$ is the work function difference between the gate and graphene.
At  equilibrium, the carrier density $n_0$ is related to the chemical potential $\mu_0$~\cite{ahcn_rmp_2009,zebrev_arxiv_2011}, in particular
at zero temperature $\mu_0= \hbar v_{\rm D}  \sqrt{\pi |n_0|}$, where $v_{\rm D}\sim 10^6~{\rm m/s}$ is the Dirac velocity. 

The voltage drop due to trapped charges depends on the trap distribution and occupation. Here we
model interface traps following References ~\cite{hooge_rps_1981,zebrev_arxiv_2011}.
Without loss of generality, we consider the presence of acceptor-like traps which are negatively charged in 
a filled state and neutral while empty. These types of traps are described here by an Anderson-type model
\begin{equation}\label{eq:Hd}
H_{{\rm t},i}=\epsilon_{i} \left(n_{(i,\uparrow)} + n_{(i,\downarrow)}\right) + U n_{(i,\uparrow)}n_{(i,\downarrow)}~,
\end{equation}
where $n_{(i, s)}$ is the number of electrons with spin $\hat{z}$-projection $s$ occupying the trap with site 
index $i$. $\epsilon_{i}$ is the trap energy in the presence of a single electron and $U$
is the electron-electron repulsion giving a finite energetic contribution in case of double trap occupancy.
Starting from the Hamiltonian of a single trap in Eq.~(\ref{eq:Hd}),
we calculate the trap occupation number of electrons at  equilibrium~\cite{grosso_book},
\begin{equation}
n_{\rm eq}(\epsilon_{i})=
\frac{2e^{(\mu_0-\epsilon_{i})/(k_{\rm B} T)}+2e^{(2\mu_0-2\epsilon_{i}-U)/(k_{\rm B} T)}}{1+2e^{(\mu_0-\epsilon_{i})/(k_{\rm B} T)}+e^{(2\mu_0-2\epsilon_{i}-U)/(k_{\rm B} T)}}~.
\end{equation}
For the sake of simplicity, here we focus on  two limiting cases: when the electrostatic repulsion is negligible ($U=0$)
\begin{equation}
n_{\rm eq}(\epsilon_{i})\Big|_{U=0}=\frac{2}{e^{(\epsilon_{i}-\mu_0)/(k_{\rm B} T)}+1}=2 n_{\rm D}(\epsilon_{i})~,
\end{equation}
and when the electrostatic repulsion is huge ($U=\infty$)
\begin{equation}
n_{\rm eq}(\epsilon_{i})\Big|_{U=\infty}=
\frac{1}{e^{[\epsilon_{i}-\mu_0-\ln(2)k_{\rm B} T]/(k_{\rm B} T)}+1}=n_{\rm D}[\epsilon_{i}-\ln(2)k_{\rm B} T]~,
\end{equation}
where $n_{\rm D}(x)=1/[e^{(x-\mu_0)/(k_{\rm B} T)}  +1]$ is the Fermi-Dirac distribution function.
By refering to the device geometry described above,  
the electric potential generated by a set of occupied traps is expressed as
\begin{eqnarray}\label{eq:Phit}
\Phi_{\rm T} ({\bf r},z,t) &=& \frac{-e}{\varepsilon_r} \int_{S} d {\bf r}^\prime \int_0^d d z^\prime  
\chi({\bf r},{\bf r}^\prime,z,z^\prime) \int_{-\Lambda}^\Lambda d \epsilon {\cal N}({\bf r}^\prime,z^\prime,\epsilon,t) ~, 
\end{eqnarray}
where $S=L_xL_y$ is the area along the plane $x$-$y$, $\Lambda$ is an energy cut-off. The kernel
\begin{eqnarray}
\chi({\bf r},{\bf r}^\prime,z,z^\prime)&=&   
\frac{1}{ \sqrt{({\bf r}-{\bf r}^\prime)^2 + (z-z^\prime)^2} }  -\frac{1}{ \sqrt{({\bf r}-{\bf r}^\prime)^2 + (z+z^\prime)^2} }~,
\end{eqnarray}
is obtained by the image charge method due to the presence of the gate metal behaving as a perfect conductor and
\begin{equation}
 {\cal N}({\bf r},z,\epsilon,t)=\sum^{N_{\rm T}}_{i=1} \rho_i({\bf r},z) \delta(\epsilon-\epsilon_i)  f_{i}(t) 
\end{equation}
is the instantaneous density of populated traps per unit volume and energy. Here
$\rho_i({\bf r},z)$ is the spatial density associated to trap  $i$ with occupation number
\begin{equation}
f_i(t)= \sum_{s=\uparrow,\downarrow} f_{(i,s)}(t)~,
\end{equation}
expressed in terms of
the occupation number resolved in spin projection $f_{(i,s)}(t)$.  $N_{\rm T}$ is the total number of traps.
We assume that a trap with negligible Coloumb repulsion, $U=0$, corresponds to a trap state delocalized along the $\hat{x}$ and $\hat{y}$ directions. Its spatial density is expressed as
\begin{equation}
 \rho_i({\bf r},z)=\frac{1}{S}\delta(z-z_i)~.
\end{equation}
Wheareas, we assume that a trap with huge Coloumb repulsion, $U=\infty$, corresponds to trap state strongly localized
and with spatial density
\begin{equation}
 \rho_i({\bf r},z)=\delta({\bf r}-{\bf r}_i)\delta(z-z_i)~.
\end{equation}
Under equilibrium conditions the population density is
\begin{equation}
{\cal N}_0({\bf r},z,\epsilon) =
\sum^{N_{\rm T}}_{i=1} \rho_i({\bf r},z) \delta(\epsilon-\epsilon_i) n_{{\rm eq}}(\epsilon_i)~,
\end{equation}
since $f_i(t)\to n_{{\rm eq}}(\epsilon_i)$.
By assuming that the traps are homogenously distribuited along the $\hat{x}$ and $\hat{y}$ directions, i.e. 
 ${\cal N}_0({\bf r},z,\epsilon)={\cal N}_0(z,\epsilon)$, one finds
\begin{eqnarray}\label{eq:PhitD2}
\Phi_{\rm T0} ({\bf r},z) &=&   \frac{ 2 \pi e}{\varepsilon_r}  \int_0^d d z^\prime  \int_{-\Lambda}^\Lambda d \epsilon 
{\cal N}_0(z^\prime,\epsilon) 
(|z-z^\prime|-|z+z^\prime| )~,
\end{eqnarray}
thus the voltage drop due to the charged traps at equilibrium reads
\begin{equation}
V_{\rm T0}= \Phi_{\rm T0} ({\bf r},0)- \Phi_{\rm T0} ({\bf r},d)=\frac{4 \pi e}{\varepsilon_r}  \int_0^d d z^\prime z^\prime  \int_{-\Lambda}^\Lambda d \epsilon \, {\cal N}_0(z^\prime,\epsilon)~.
\end{equation}

\section{Localized/delocalized trap density fluctuations}
\label{sec:RT1}

Charge density noise arises from fluctuations of the density of populated trap states 
per unit volume and energy ${\cal N}_{{\rm T}}$ around its average value ${\cal N}_{{\rm T}0}$.
In this Section we introduce a theoretical approach to describe these fluctuations based on the McWorther model~\cite{mcwhorther_1957}. 
We assume that charge fluctuations in the substrate traps are due to  trapping-recombination processes~\cite{kogan_book}.
Since the time between two successive transitions is usually many times longer than the time of relaxation 
(equilibration) of the crystal after each such transition,  trapping and recombinations can be considered 
as discrete Markov processes~\cite{kogan_book}.

First of all, we consider the case with zero repulsion, $U=0$.
The occupancy number $X_{(i,s)}$ of the trap state labeled by the site index $i$ and the spin projection $s$ is a random variable, 
we consider that each trap state can be empty ($X_{(i,s)}=0$) or occupied by a single electron ($X_{(i,s)}=1$), 
and it randomly switches between these states (excluding spin-flip processes) with time-independent rates (stationary process)\cite{lax_rmp_1960}. 
The conditional probability that the trap state with site index and spin projection $(i,s)$ at time $t$ has the occupation number $X_{(i,s)}$  if 
the trap state with site index and spin projection  $(j,s^\prime)$ at time $t_0$ has the occupation number $X_{(j,s^\prime)}$ is written as
\begin{eqnarray}\label{eq:conditional}
 P[X_{(i,s)}(t)|X_{(j,s^\prime)}(t_0)]&=& \delta_{i, j} \delta_{s, s^\prime} P[X_{(i,s)}( t-t_0)|X_{(i,s)}( 0)]~,
\end{eqnarray}
\begin{equation}
P[X_{(i,s)}( t)|X_{(i,s)}( 0)]= p[X_{(i,s)}( t)|X_{(i,s)}( 0)] +w(X_{(i,s)})~,
\end{equation}
where $w(X_{(i,s)})$ is the stationary probability which can take the two values
$w(1_{(i,s)})=f_{(i,s)}$ and $w(0_{(i,s)})=1-f_{(i,s)}$, and the matrix $p$ has the form
\begin{eqnarray}
p[X_{(i,s)}( t)|X_{(i,s)}( 0)] &=& 
\left( \matrix{ p[0_{(i,s)}( t)|0_{(i,s)}( 0)] & p[0_{(i,s)}( t)|1_{(i,s)}( 0)] \cr
p[1_{(i,s)}( t)|0_{(i,s)}( 0)]  & p[1_{(i,s)}( t)|1_{(i,s)}( 0)] \cr} \right)~.
\end{eqnarray}
The Kronecker delta appears in Eq.~(\ref{eq:conditional}) because different trap states are uncorrelated.
Moreover, we assume that the stationary probability coincides with the equilibrium occupation function, i.e. $f_{(i,s)}=n_{\rm D}(\epsilon_{i})$.
The matrix $p$ is solution of the general Kolmogorov equation~\cite{kogan_book}
\begin{equation}
\frac{d p}{d t}=-M_{(i,s)} \cdot p~
\end{equation}
where we consider regimes where the rate matrix does not depend on the electron spin 
\begin{eqnarray}
M_{(i,s)} = \left( \matrix{ \lambda_{i,00} & -\lambda_{i,11} \cr
-\lambda_{i,00}  & \lambda_{i,11} \cr} \right)~,
\end{eqnarray}
and we choose equilibrium initial conditions
\begin{eqnarray}
p[X_{(i,s)}( 0)|X_{(i,s)}( 0)]&=& 
\left( \matrix{ 1-w(0_{(i,s)}) & -w(0_{(i,s)}) \cr
-w(1_{(i,s)}) & 1-w(1_{(i,s)}) \cr} \right)~.
\end{eqnarray}
The solution of the Kolmogorov equation is the following matrix
\begin{equation}\label{eq:p}
p[X_{(i,s)}( t)|X_{(i,s)}( 0)]=
\left( \matrix{ w(1_{(i,s)}) & -w(0_{(i,s)}) \cr
-w(1_{(i,s)})  & w(0_{(i,s)}) \cr} \right)
 e^{-\gamma_i t}~,
 \end{equation}
where $\gamma_i=\lambda_{i,00} +\lambda_{i,11} $ is the switching rate between the two states
of the stochastic process. In the following Section we will specify its dependence on the trap position
in the substrate.

Now, we consider the limiting case of infinite repulsion, $U=\infty$. Since double occupancy of each
trap state by two electrons with opposite spin orientations is energetically forbidden, the relevant
stochastic variable taking values $0$ and $1$ is  
\begin{equation}
X_i= X_{(i,\uparrow)}+ X_{(i,\downarrow)}~,
\end{equation}
and the conditional probability that the trap with site index $i$ at time $t$ has the occupation number $X_i$  if 
the trap with site index $j$ at time $t_0$ has the occupation number $X_j$ is written as
\begin{eqnarray}\label{eq:conditional_U}
 P[X_i(t)|X_j(t_0)]&=& \delta_{i, j}  P[X_i( t-t_0)|X_i( 0)]~,
\end{eqnarray}
\begin{equation}
P[X_i( t)|X_i( 0)]= q[X_i( t)|X_i( 0)] +v(X_{i})
\end{equation}
where $v(1_i)=f_{i}$ and $v(0_i)=1-f_{i}$, and $q$ has the following matrix form
\begin{eqnarray}
q[X_i( t)|X_i( 0)] &=& 
\left( \matrix{ q[0_i(t)|0_i(t)] & q[0_i (t)|1_i(t)] \cr
q[1_i(t)|0_i(t)] & q[1_i(t)|1_i(t)] \cr} \right)~,
\end{eqnarray}
the Kronecker delta appears in Eq.~(\ref{eq:conditional_U}) because  different trap sites are uncorrelated.
Similarly to the previous case, by solving the Kolmogorov equation we obtain the following matrix
\begin{equation}\label{eq:q}
q[X_{(i,s)}( t)|X_{(i,s)}( 0)]=
\left( \matrix{ v(1_{i}) & -v( 0_{i}) \cr
-v(1_{i})  & v(0_{i}) \cr} \right)
 e^{-\gamma_i t}~,
 \end{equation}
and we assume that the stationary probability corresponds to the equilibrium occupation function, i.e. 
$f_{i}=n_{\rm D}[\epsilon_{i}-\ln(2)k_{\rm B} T]$.

In order to describe the charge density noise, we analyze the fluctuations of the density of populated trap states 
per unit volume and energy ${\cal N}_{{\rm T}}$ around its average value ${\cal N}_{{\rm T}0}$, which is induced by the trapping-recombination process 
and it is written as
\begin{equation}
{\cal N}_{\rm T}({\bf r},z,\epsilon,t)=
{\cal N}_{{\rm T},0}({\bf r},z,\epsilon)+\delta {\cal N}_{\rm T}({\bf r},z,\epsilon,t)~,
\end{equation}
where
\begin{equation}\label{eq:nt_random}
\delta {\cal N}_{\rm T}({\bf r},z,\epsilon,t) =  \sum^{N_{\rm T}}_{i=1} \rho_i({\bf r},z) \delta(\epsilon-\epsilon_i) \Delta_{i}(t)~,
\end{equation}
and
\begin{equation}
 \Delta_{i}(t) = \sum_{s=\uparrow,\downarrow}  X_{(i,s)}(t) - n_{\rm eq}(\epsilon_i)
\end{equation}
is the deviation of the occupancy number for the trap with site index $i$ with respect to its value at equilibrium.
By using the conditional probabilities derived in this Section, we obtain for both $U=0$ and $U=\infty$ the average value
\begin{equation}
\langle  \Delta_{i}(t) \rangle =0~.
\end{equation}
The correlators instead depend on the localized/delocalized assumpion on the trap distribution.
In the absence of repulsion, $U=0$, we have 
\begin{eqnarray}
\langle \Delta_{i}(t) \Delta_{j}(t)  \rangle =
2 \delta_{ij} [1-n_{\rm D}(\epsilon_{i})]n_{\rm D}(\epsilon_{i})e^{-\gamma_{i}t}~,
\end{eqnarray}
instead, for infinite repulsion, $U=\infty$, 
\begin{eqnarray}
\langle \Delta_{i}(t) \Delta_{j}(t)  \rangle = &=& 
\delta_{ij}   [1-n_{\rm D}(\epsilon_{i}-\ln(2)k_{\rm B}T)]
n_{\rm D}(\epsilon_{i}-\ln(2)k_{\rm B}T)e^{-\gamma_{i}t}~.
\end{eqnarray}

\section{Charge carrier density noise}

\begin{figure}[ht]
\centering
\includegraphics[width=.49\columnwidth]{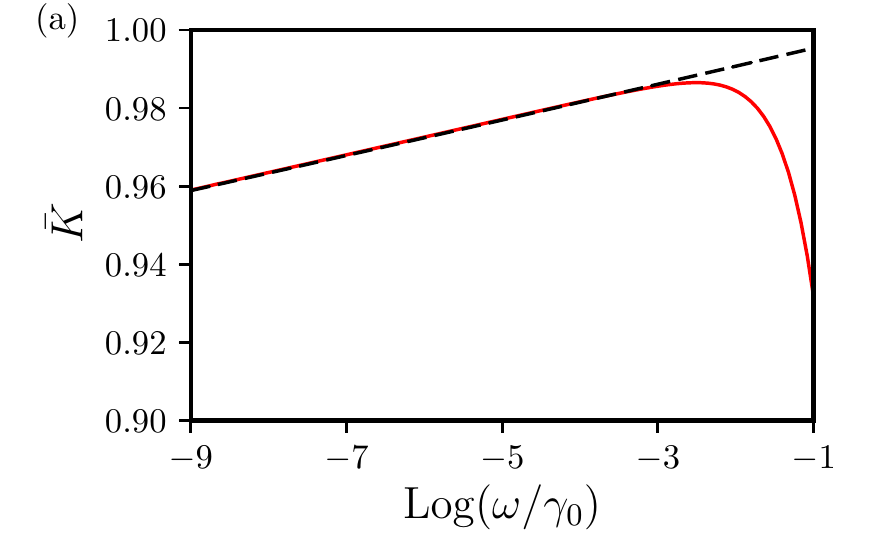}
\includegraphics[width=.49\columnwidth]{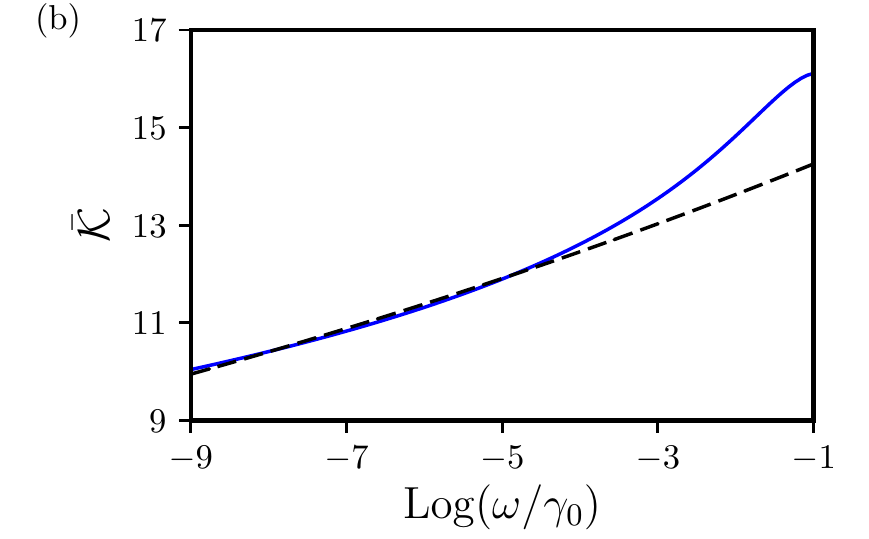}
\caption{Panel(a), $U=0$: $\bar{K}$ (red solid line), defined in Eq.~(\ref{eq:K}), as a function of $\omega/ \gamma_0$ in a logarithmic scale,
compared with the function $\omega^\beta$ (black dashed line), where $\beta=2.0 \times 10^{-3}$.
Panel(b), $U=\infty$: $\bar{\cal K}$ (blue solid line), defined in Eq.~(\ref{eq:Kcal}), as a function of $\omega /\gamma_0$ in a logarithmic scale,
compared with the function $A \omega^\beta$ (black dashed line), where $\beta=1.9 \times 10^{-2}$ and $A=14.9$.
In both panels we set $\ell_0/d=10^{-3}$.
}
\label{fig:Kfunctions}
\end{figure}

Since the time scale of fluctuations of carriers in graphene is much shorter than the time scale of the charge traps fluctuations~\cite{kogan_book}, 
we assume that  charge carriers in graphene responds instantaneously to the fluctuations of the trapped carriers $\delta {\cal N}_{\rm T}(t)$. 
Here, we focus on the high doping regime ($n_0 \sim 10^{12}~{\rm cm}^{-2}$), so we neglect the quantum corrections to the classical capacitance~\cite{fernandez_prb_2007}.
Under these conditions, by keeping fixed $V_{\rm G}$ and by expanding Eq.~(\ref{eq:Vg}) around the equilibrium values, one has
\begin{equation}\label{eq:dn}
\delta n(t)=\frac{C_{\rm g}}{-e} \delta V_{\rm T}(t)~,
\end{equation}
where $\delta n(t)=n(t)-n_0$ and $ \delta V_{\rm T}(t)=V_{\rm T}(t)-V_{\rm T0}$.
The relation above links the fluctuations of charge carrier density  with the trapping-recombination processes, 
and it allows to write the charge carrier density power spectrum in terms of the fluctuations of charges in the 
substrate traps as
\begin{eqnarray}\label{eq:Sn_def}
{\cal S}_{ n} (\omega)&= &\int_0^\infty \frac{d t}{\pi}\cos(\omega t) \langle \delta n(t) \delta   n(0) \rangle~,\\
&=&\frac{C_{\rm g}^2}{e^2 } \int_0^\infty \frac{d t}{\pi}\cos(\omega t) \langle \delta  V_{\rm T}(t)  \delta V_{\rm T}(0) \rangle. \nonumber
\end{eqnarray}
The correlator of the voltage fluctuations is linearly related to the fluctuations of the density of populated traps 
per unit volume and energy which read, for both cases $U=0$ and $U=\infty$
\begin{eqnarray}
&&
\langle \delta V_{\rm T} (t) \delta V_{\rm T} (0) \rangle = \frac{e^2}{\varepsilon_r^2} 
\int_{S} d {\bf r}^\prime \int_0^d d z^\prime  \int_{S} d {\bf r}^{\prime\prime} \int_0^d d z^{\prime\prime}  
\int_{-\Lambda}^\Lambda d \epsilon^\prime \int_{-\Lambda}^\Lambda d \epsilon^{\prime\prime} 
\\
&\times&  \chi({\bf r},{\bf r}^\prime,d,z^\prime)  \chi({\bf r},{\bf r}^{\prime\prime},d,z^{\prime\prime}) 
\langle \delta {\cal N}({\bf r}^\prime,z^\prime,\epsilon^\prime,t) \delta {\cal N}({\bf r}^{\prime\prime},z^{\prime\prime},\epsilon^{\prime\prime},0) \rangle ~, \nonumber
\end{eqnarray}
where
\begin{eqnarray}
\langle \delta  {\cal N}({\bf r}^\prime,z^\prime,\epsilon^\prime,t) \delta {\cal N}({\bf r},z,\epsilon,0)   \rangle
&=& \sum^{N_{\rm T}}_{i=1} \rho_i({\bf r}^\prime,z^\prime)  \delta(\epsilon^{\prime}-\epsilon_i) \rho_i({\bf r},z) \delta(\epsilon-\epsilon_i) \\
&\times&
\langle \Delta_{i}(t) \Delta_i(0) \rangle ~. \nonumber
\end{eqnarray}
For delocalized traps, i.e. in the absence of repulsion $U=0$,  we have
\begin{eqnarray}\label{eq:nn}
\langle \delta {\cal N}({\bf r}^\prime,z^\prime,\epsilon^\prime,t) \delta  {\cal N}({\bf r},z,\epsilon,0)  \rangle
&=&   \frac{2}{S} \delta(z^\prime-z) \delta(\epsilon^\prime-\epsilon)   
\sum^{N_{\rm T}}_{i=1} 
 \rho_i({\bf r},z)  \\
&\times& \delta(\epsilon-\epsilon_i)  [1-n_{\rm D}(\epsilon_{i})]n_{\rm D}(\epsilon_{i})e^{-\gamma_{i}t}~.\nonumber
\end{eqnarray}
Following the McWorther model, we consider the 
switching rates $\gamma_i $ depending on the trap position along the direction perpendicular to the graphene layer
as follows
\begin{equation}\label{eq:gamma}
 \gamma_i=\gamma(z_i)=\gamma_0 e^{-|z_i-d|/\ell_0}~,
\end{equation}
where typical orders of magnitude of the tunneling parameters are $\gamma_0  \sim 10^{10}~{\rm s}^{-1}$ and $\ell_0  \sim 1~$\AA, respectively~\cite{balandin_natnano_2013}. 
The density of trap states is expressed as
\begin{equation}\label{eq:D}
D(\epsilon)=\frac{2}{S}  \sum^{N_{\rm T}}_{i=1} \delta(z-z_i) \delta(\epsilon-\epsilon_i)~, 
\end{equation}
here we consider that the traps are distribuited uniformly in the substrate, so that
$D(\epsilon)$ can be treated as a function only of energy. We remark  that the factor two in Eq.~(\ref{eq:D}) is 
due to the spin degeneracy.
By using Eq.~(\ref{eq:D}), one has the compact form
\begin{eqnarray}\label{eq:PP_0}
\langle \delta V_{\rm T} (t) \delta V_{\rm T} (0) \rangle &=& \left(\frac{4 \pi d e}{\varepsilon_r}\right)^2
\frac{1}{S} FK(t)~,
\end{eqnarray}
where 
\begin{equation}
F =  \int_{-\Lambda}^\Lambda d \epsilon  D(\epsilon) [1-n_{\rm D}(\epsilon)]n_{\rm D}(\epsilon)~,  
\end{equation}
and
\begin{equation}
K(t) =  \int_0^d dz \left( \frac{z}{d} \right)^2 e^{-\gamma(z)t}~.
\end{equation}
By replacing Eq.~({\ref{eq:PP_0}}) in Eq.~(\ref{eq:Sn_def}), one has
\begin{equation}\label{eq:Sn_0}
 {\cal S}_{ n} (\omega)= \frac{F}{S}  \frac{ \ell_0}{2 \omega} \bar{K}(\omega)~,
\end{equation}
where we have used
\begin{equation}
\int_0^\infty \frac{d t}{\pi}\cos(\omega t)  K(t) =
\frac{ \ell_0}{2 \omega} \bar{K}(\omega)~,
\end{equation}
and
\begin{equation}\label{eq:K}
\bar{K}(\omega)=\frac{2}{\pi }\int_{\gamma_0 e^{-d/\ell_0}/\omega}^{{\gamma_0/\omega}}  
 d \xi [1+(\ell_0/d) \ln(\xi \omega/\gamma_0)]^2    \frac{1}{\xi^2+1}~.
\end{equation}
The dimensionless function $\bar{K}(\omega)$ gives the deviation of the frequency dependence of the power spectrum from the usual $1/\omega$ law. 
Fig.~\ref{fig:Kfunctions}~(a) shows $\bar{K}(\omega)$ (red solid line) as a function of frequency 
and it is compared with the function $\omega^\beta$ with $\beta=2 \times 10^{-3}$ (black dashed line), 
having fixed $\ell_0/d=10^{-3}$.
The comparison shown in Fig.~\ref{fig:Kfunctions}~(a) proves that the charge carrier density behaves as $1/f^{0.998} \sim 1/f$ for sufficiently small frequencies.
In the power spectrum expressed in Eq~(\ref{eq:Sn_0}) the temperature dependence is exclusively in the term $F$. 
By assuming  that density of trap states is a smooth function, 
and considering that $[1-n_{\rm D}(\epsilon)]n_{\rm D}(\epsilon)$ at low temperatures $k_{\rm B} T \ll \mu_0$
is peaked at $\epsilon = \mu_0$, one can approximate the $F$ term as
\begin{eqnarray}\label{eq:F}
F &\approx&  \int_{-\Lambda}^\Lambda d \epsilon  \left[D(\mu_0)+ \frac{d D(\epsilon)}{d \epsilon}\Bigg|_{\epsilon=\mu_0} (\epsilon-\mu_0) +
 \frac{d^2 D(\epsilon)}{d \epsilon^2}\Bigg|_{\epsilon=\mu_0} \frac{(\epsilon-\mu_0)^2}{2} 
\right ]\\
&\times&[1-n_{\rm D}(\epsilon)]n_{\rm D}(\epsilon)=D(\mu_0) k_{\rm B} T + \frac{\pi^2}{6} \frac{d^2 D(\epsilon)}{d \epsilon^2}\Bigg|_{\epsilon=\mu_0} (k_{\rm B} T)^3
~,  \nonumber
\end{eqnarray}

In the case of strong repulsion we have 
\begin{eqnarray}\label{eq:nn_U}
 && 
\langle \delta {\cal N}({\bf r}^\prime,z^\prime,\epsilon^\prime,t)\delta  {\cal N}({\bf r},z,\epsilon,0)  \rangle
=  \delta({\bf r}^\prime -{\bf r})  \delta(z^\prime-z) \delta(\epsilon^\prime-\epsilon)   
\sum^{N_{\rm T}}_{i=1} 
 \rho_i({\bf r},z)  \nonumber  \\
&\times&   \delta(\epsilon-\epsilon_i)  [1-n_{\rm D}(\epsilon_{i}-\ln(2) k_{\rm B} T)]n_{\rm D}(\epsilon_{i}-\ln(2) k_{\rm B} T)
e^{-\gamma_{i}t}~,
\end{eqnarray}
where we put again $ \gamma_i=\gamma_0 e^{-|z_i-d|/\ell_0}$.
Here, the density of trap states is expressed as
\begin{equation}
{\cal D}(\epsilon)=  2\sum^{N_{\rm T}}_{i=1} \delta ({\bf r}-{\bf r}_i) \delta(z-z_i) \delta(\epsilon-\epsilon_i)~.
\end{equation}
For traps distribuited uniformly in the substrate, the density of trap states only depends on the corresponding
energy and one has
\begin{eqnarray}\label{eq:PP_U}
\langle \delta V_{\rm T} (t) \delta V_{\rm T} (0) \rangle &=& \left(\frac{  e}{\varepsilon_r}\right)^2 
{\cal K}(t) {\cal F}~,
\end{eqnarray}
where
\begin{equation}
{\cal F} =  \int_{-\Lambda}^\Lambda d \epsilon   {\cal D}(\epsilon)  [1-n_{\rm D}(\epsilon_{i}-\ln(2) k_{\rm B} T)]n_{\rm D}(\epsilon_{i}-\ln(2) k_{\rm B} T)~,
\end{equation}
and
\begin{equation}\label{eq:Kcal_t}
{\cal K}(t) =  -\pi \int_0^d dz\ln\left[1- \left( \frac{z}{d} \right)^2\right] e^{-\gamma(z)t}~.
\end{equation}
By replacing Eq.~({\ref{eq:PP_U}}) in Eq.~(\ref{eq:Sn_def}), the power spectrum takes the form
\begin{equation}\label{eq:Sn_U}
 {\cal S}_{ n} (\omega)= \frac{\cal F}{(4 \pi d)^2}  \frac{ \ell_0}{2 \omega} \bar{\cal K}(\omega)~,
\end{equation}
where we have used
\begin{equation}
\int_0^\infty \frac{d t}{\pi}\cos(\omega t)  {\cal K}(t) =
\frac{\ell_0}{2 \omega} \bar{{\cal K}}(\omega)~,
\end{equation}
and
\begin{equation}\label{eq:Kcal}
\bar{\cal K}(\omega)=-2\int_{\gamma_0 e^{-d/\ell_0}/\omega}^{{\gamma_0/\omega}}  
 d \xi \ln\{1-[1+(\ell_0/d) \ln(\xi \omega/\gamma_0)]^2\}    \frac{1}{\xi^2+1}~.
\end{equation}
The dimensionless function $\bar{\cal K}(\omega)$ gives the deviation of the frequency dependence of the power spectrum from the usual $1/\omega$.
Fig.~\ref{fig:Kfunctions}~(b) shows $\bar{\cal K}(\omega)$ (blue solid line) as a function of frequency 
and it is compared with the function $A \omega^\beta$ with $\beta=1.9 \times 10^{-2}$ and $A=14.9$ (black dashed line),
for $\ell_0/ d=10^{-3}$.
From the comparison shown in Fig.~\ref{fig:Kfunctions}~(b), it is clear that also for large electrostatic repulsion
the charge carrier density noise behaves as $1/f^{0.98}$  at low frequencies.
From the analysis in Fig.~\ref{fig:Kfunctions} we note that
the deviation of frequency dependence of the power spectrum ${\cal S}_n$ from the $1/\omega$ law 
is larger for localized traps ($U=\infty$), see panel~(b), 
with respect to the case with delocalized traps ($U=0$), see panel~(a).
In the power spectrum expressed in Eq.~(\ref{eq:Sn_U}),
the term ${\cal F}$ contains the temperature dependence of the charge carrier density noise in the case with $U=\infty$. 
 Similarly to the case with delocalized traps,  we consider the low temperature limit ($k_{\rm B} T \ll \mu_0$) and
a smooth density of trap states. Under these conditions one can approximate $\cal F$ as
\begin{eqnarray}\label{eq:Fcal}
{\cal F} &\approx& {\cal D}(\mu_0) k_{\rm B} T + 
\ln(2) \frac{d {\cal D}(\epsilon)}{d \epsilon}\Bigg|_{\epsilon=\mu_0} (k_{\rm B} T)^2\\
&+&\frac{\pi^2+3\ln^2(2)}{6}
 \frac{d^2 {\cal D}(\epsilon)}{d \epsilon^2}\Bigg|_{\epsilon=\mu_0}
(k_{\rm B} T)^3
\nonumber ~.
\end{eqnarray}
Thus the first correction to the linear temperature dependence is a quadratic, differently from the case with $U=0$
where the first correction to the linear temperature dependence is  cubic, see Eq.~(\ref{eq:F}).

\section{Conclusions}

We have studied the charge density noise in graphene by using a minimal approach based on the McWhorter model~\cite{mcwhorther_1957} in the large doping regime  ($n\sim 10^{12}~{\rm cm}^{-2}$). 
According to this description, the charge density noise is induced by the charge fluctuations in the electron traps within the substrate. 
Here, we have analyzed two types of electron traps, namely spatially localized and delocalized traps. Within our model, we have shown how the nature of  the electron traps affects  
the temperature dependence of the charge carriers noise. 
In particular, by assuming that the density of trap states is a
smooth function of energy then single/double spin occupancy of trap states (corresponding to localized/delocalized traps) influences the temperature 
dependence of noise amplitude only to second order. Moreover, we have analyzed how the nature of the electron traps modifies the usual $1/f$ low-frequency dependence 
of the charge noise. In terms of a phenomenological formula $1/f^\alpha$ for the power spectrum, 
we have found that the deviation $\beta=1-\alpha$ is one order of magnitude larger for localized traps ($\beta \sim 10^{-2}$) 
with respect to  delocalized traps ($\beta \sim 10^{-3}$).
 Finally, we remark that our analysis, in the large doping regime, applies to any two-dimensional electron system. 
Future work will be devoted to analyze low frequency charge density noise in proximity of the charge neutrality point
where the linear dispersion of the graphene monolayer may induce peculiar effects.

\end{document}